\documentclass[12pt]{article} 

\usepackage{amsmath,amsfonts,amssymb,latexsym,cite}

\newcounter{thMM}
\newcounter{leMM}
\newcounter{deFF}
\newcounter{exMP}
\title{\normalsize\bf 
PRECANONICAL QUANTIZATION OF YANG-MILLS FIELDS AND 
THE FUNCTIONAL SCHR\"ODINGER REPRESENTATION
\\
}
\author{ 
Igor V. Kanatchikov\thanks{ {\; }On leave from 
Tallinn Technical University, Tallinn, Estonia.}  
\thanks{ {\; }{\em E-mail:\/ }ikanat@physik.fu-berlin.de. } \\ 
Institut f\"ur Theoretische Physik,  
Freie Universit\"at  Berlin, \\ 
Arnimallee 14, D-14195 Berlin,  Germany\\ 
}

\catcode `\@=11
\@addtoreset{equation}{section}

\def\section{\@startsection {section}{1}{\z@}{-3.5ex plus -1ex minus
     -.2ex}{2.3ex plus .2ex}{\normalsize\bf}}
\def\subsection{\@startsection{subsection}{2}{\z@}{-3.25ex plus -1ex minus
 -.2ex}{1.5ex plus .2ex}{\normalsize\bf}}

\def\thebibliography#1{\section*{References}
\small\rm\list
 {[\arabic{enumi}]}{\settowidth\labelwidth{[#1]}\leftmargin\labelwidth
 \advance\leftmargin\labelsep\usecounter{enumi}}
 \def\newblock{\hskip .11em plus .33em minus -.07em}
 \sloppy\clubpenalty4000\widowpenalty4000
 \sfcode`\.=1000\relax}

\catcode `\@=12 

\begin{document}

\date{}

\maketitle

\begin{abstract}
Precanonical quantization of pure Yang-Mills fields,   
which is based on the covariant De~Donder-Weyl (DW) Hamiltonian 
formulation, and its connection 
with the functional Schr\"odinger representation 
in the temporal gauge are discussed. 
The mass gap problem is related to a  
finite dimensional spectral problem 
for a generalized Clifford-valued magnetic Schr\"odinger operator   
which represents the DW Hamiltonian operator.  
\end{abstract} 



\newcommand{\beq}{\begin{equation}}
\newcommand{\eeq}{\end{equation}}
\newcommand{\beqa}{\begin{eqnarray}}
\newcommand{\eeqa}{\end{eqnarray}}
\newcommand{\nn}{\nonumber}

\newcommand{\half}{\frac{1}{2}}

\newcommand{\xt}{\tilde{X}}

\newcommand{\uind}[2]{^{#1_1 \, ... \, #1_{#2}} }
\newcommand{\lind}[2]{_{#1_1 \, ... \, #1_{#2}} }
\newcommand{\com}[2]{[#1,#2]_{-}} 
\newcommand{\acom}[2]{[#1,#2]_{+}} 
\newcommand{\compm}[2]{[#1,#2]_{\pm}}

\newcommand{\lie}[1]{\pounds_{#1}}
\newcommand{\co}{\circ}
\newcommand{\sgn}[1]{(-1)^{#1}}
\newcommand{\lbr}[2]{ [ \hspace*{-1.5pt} [ #1 , #2 ] \hspace*{-1.5pt} ] }
\newcommand{\lbrpm}[2]{ [ \hspace*{-1.5pt} [ #1 , #2 ] \hspace*{-1.5pt}
 ]_{\pm} }
\newcommand{\lbrp}[2]{ [ \hspace*{-1.5pt} [ #1 , #2 ] \hspace*{-1.5pt} ]_+ }
\newcommand{\lbrm}[2]{ [ \hspace*{-1.5pt} [ #1 , #2 ] \hspace*{-1.5pt} ]_- }

\newcommand{\pbr}[2]{ \{ \hspace*{-2.2pt} [ #1 , #2\hspace*{1.4 pt} ] 
\hspace*{-2.3pt} \} }
\newcommand{\nbr}[2]{ [ \hspace*{-1.5pt} [ #1 , #2 \hspace*{0.pt} ] 
\hspace*{-1.3pt} ] }

\newcommand{\we}{\wedge}
\newcommand{\nbrpq}[2]{\nbr{\xxi{#1}{1}}{\xxi{#2}{2}}}
\newcommand{\lieni}[2]{$\pounds$${}_{\stackrel{#1}{X}_{#2}}$  }

\newcommand{\rbox}[2]{\raisebox{#1}{#2}}
\newcommand{\xx}[1]{\raisebox{1pt}{$\stackrel{#1}{X}$}}
\newcommand{\xxi}[2]{\raisebox{1pt}{$\stackrel{#1}{X}$$_{#2}$}}
\newcommand{\ff}[1]{\raisebox{1pt}{$\stackrel{#1}{F}$}}
\newcommand{\dd}[1]{\raisebox{1pt}{$\stackrel{#1}{D}$}}
\newcommand{\der}{\partial}
\newcommand{\oo}{$\Omega$}
\newcommand{\Om}{\Omega}
\newcommand{\om}{\omega}
\newcommand{\eps}{\epsilon}
\newcommand{\si}{\sigma}
\newcommand{\Lm}{\bigwedge^*}

\newcommand{\inn}{\hspace*{2pt}\raisebox{-1pt}{\rule{6pt}{.3pt}\hspace*
{0pt}\rule{.3pt}{8pt}\hspace*{3pt}}}
\newcommand{\sro}{Schr\"{o}dinger\ }
\newcommand{\bm}{\boldmath}
\newcommand{\vol}{\omega}
               \newcommand{\dvol}[1]{\der_{#1}\inn \vol}

\newcommand{\bd}{\mbox{\bf d}}
\newcommand{\bder}{\mbox{\bm $\der$}}
\newcommand{\bI}{\mbox{\bm $I$}}

\newcommand{\be}{\beta} 
\newcommand{\ga}{\gamma} 
\newcommand{\de}{\delta} 
\newcommand{\Ga}{\Gamma} 
\newcommand{\gmu}{\gamma^\mu}
\newcommand{\gnu}{\gamma^\nu}
\newcommand{\ka}{\kappa}
\newcommand{\hka}{\hbar \kappa}
\newcommand{\al}{\alpha}
\newcommand{\lapl}{\bigtriangleup}
\newcommand{\psib}{\overline{\psi}}
\newcommand{\Psib}{\overline{\Psi}}
\newcommand{\Phib}{\overline{\Phi}}
\newcommand{\derts}{\stackrel{\leftrightarrow}{\der}}
\newcommand{\what}[1]{\widehat{#1}}

\newcommand{\bx}{{\bf x}}
\newcommand{\bk}{{\bf k}}
\newcommand{\bq}{{\bf q}}

\newcommand{\omk}{\omega_{\bf k}} 
\newcommand{\lpl}{\ell}
\newcommand{\zb}{\overline{z}} 

\newcommand{\dv}{\mbox{\sf d}}

\newcommand{\BPsi}{{\bf \Psi}} 
\newcommand{\BH}{{\bf H}} 
\newcommand{\BS}{{\bf S}} 
\newcommand{\BN}{{\bf N}} 


\section{Introduction}

The problem of quantum Yang-Mills 
theory has been extensively discussed 
for more than three decades resulting in such breakthroughs as the 
asymptotic freedom, the renormalizability proof, 
the Faddeev-Popov technique, the BRST symmetry, the Seiberg-Witten theory,   
the duality and others.  
However, some of the fundamental issues such as 
the mathematically satisfactory definition of quantum Yang-Mills theory,  
the confinement problem and the existence of the mass gap 
are still not properly understood. 
It is therefore desirable to explore the potential of new approaches 
in solving these problems. 

A new approach of {\em precanonical\/} quantization of fields which is 
based on a manifestly covariant (space-time symmetric) version 
of the Hamiltonian formalism in field theory 
\cite{gimm,romp98,bial96,paufler2002,roman,
sardan,norris,deleon,helein,dw-refs} 
has been proposed recently in our papers 
\cite{qs96,bial97,lodz98,ijtp2001,opava2001}. 
It is conceptually different from the standard picture of quantum 
field theory as an infinite dimensional quantum mechanics. 
Instead, precanonical quantization is based on the representation  
of classical fields as multi-parameter generalized Hamiltonian systems 
in the sense that all space-time variables 
enter on  equal footing as analogues of the time 
variable in 
Hamiltonian mechanics. This guarantees the 
manifest covariance of the formulation. The corresponding 
analogue of the configuration  space is a bundle of field variables over 
the space-time; the classical field configurations are 
sections of this bundle.  
The wave functions of quantum fields are functions 
on this bundle, not functionals of its sections. 
Quantization of the abovementioned  
generalized Hamiltonian systems representing field theories 
leads to a multi-parameter, Clifford-algebraic generalization 
of quantum mechanics to field theory which reduces to the familiar 
complex Hilbert space quantum mechanics 
in the case of (0+1)--dimensional space-time whose corresponding 
Clifford algebra is the algebra of the complex numbers. 
 
The manifest covariance of the formulation 
and the finite dimensionality of the 
analogue of the configuration space
are the obvious advantages of the precanonical approach 
to field quantization.  
However, the relations of this description of quantum fields 
to the standard techniques 
and notions of quantum field theory are not yet fully understood.  
An  exception is a connection between precanonical 
quantization and the functional Schr\"odinger representation 
found in \cite{pla2001}. 

In this paper our aim is to consider precanonical quantization 
of pure Yang-Mills theory 
 and to demonstrate its connection with 
the functional Schr\"odinger representation 
of quantum Yang-Mills theory. As a by-product, 
the precanonical approach is argued to relate the 
mass gap problem 
in pure Yang-Mills theory 
to a finite dimensional spectral problem 
for a generalized (Clifford-valued) magnetic Schr\"odinger operator 
in the space of gauge potentials.

\section{The De Donder-Weyl Hamiltonian formulation of Yang-Mills theory }

The Lagrangian density of pure Yang-Mills theory is given by 
\beq
L= - \frac{1}{4} F_{a\mu\nu}F^{a\mu\nu}, 
\eeq
where 
\beq
F^a_{\mu\nu} := \der_\mu A^a_\nu - \der_\nu A^a_\mu 
+ g C^a{}_{bc} A^b_\mu A^c_\nu ,  
\eeq 
$g$ is the Yang-Mills  self-coupling constant and 
$C_{abc}$ are totally antisymmetric structure constants which 
fulfill the Jacobi identity  
$$
C^e{}_{ab}C^d{}_{ec} + C^e{}_{bc}C^d{}_{ea} + C^e{}_{ca}C^d{}_{eb}=0.
$$


Following the De Donder-Weyl (DW) Hamiltonian formulation 
\cite{gimm,romp98,bial96,paufler2002,roman,sardan,dw-refs} 
we define the polymomenta
\beq
\pi_a^{\nu\mu} 
:= \frac{\der L}{\der(\der_\mu A^a_\nu)} 
= -\der^\mu A_a^\nu + \der^\nu A_a^\mu - g C_a{}_{bc} A^b_\mu A^c_\nu 
= - F_a^{\mu\nu}, 
\eeq
and the 
DW Hamiltonian 
\beq
H
= 
\pi_a^{\nu\mu}\der_\mu A^a_\nu - L 
= 
-\frac{1}{4} \pi_{a\mu\nu} \pi^{a\mu\nu} 
+ \frac{g}{2} C^a{}_{bc}A^b_\mu A^c_\nu \pi_a^{\mu\nu} . 
\eeq
Then the Yang-Mills field equations take the DW Hamiltonian form: 
\beqa
\der_\mu \pi^{\nu\mu}_a &=& -\frac{\der H}{\der A^a_\nu} 
\;\;=\;\; 
-g\, C_{abc}A^b_\mu\pi^{\nu\mu}_c  ,
 \\ 
\der_{[\mu} A^a_{\nu]} &=& \, \frac{\der H}{\der \pi^{\nu\mu}_a}  
\quad\! =\;\; \frac{1}{2} \pi_{\mu\nu}^a - 
\frac{1}{2}g\, C^a_{bc}A^b_\mu A^c_\nu .
\eeqa

The antisymmetrization in the left hand side of the 
second equation makes the  DW Hamiltonian equations 
consistent with the primary constraints 
\beq
\pi_a^{\mu\nu}+\pi_a^{\nu\mu} \approx 0  
\eeq
which follow from (2.3). 
It ensures the gauge invariance of (2.6). 
Note that instead of attempting to generalize the techniques of 
the constrained dynamics to the DW Hamiltonian formulation, 
in this paper the constraints (2.7) are taken into account 
heuristically. 
 

An earlier consideration  of classical Yang-Mills fields  
using the closely related multisymplectic framework 
can be found in \cite{kondracki}. 
After the present paper has been accepted for publication the paper 
by L\'opez and Marsden has appeared \cite{marco} which studies 
geometrical issues related to the constrained dynamics for 
the DW Hamiltonian formulation of gauge theories.  
 
\section{Precanonical quantization of Yang-Mills theory}

The Poisson-Gerstenhaber brackets underlying precanonical quantization 
of this section,  which also enable us to represent the 
DW Hamiltonian formulation of Yang-Mills  fields, eqs. (2.5), (2.6),  
in Poisson bracket form,  can be obtained from the 
reduced polysymplectic form  (c.f. \cite{marco}) 
\beq
\Omega := d \pi_a^{\mu\nu}\we dA_{[\mu}^a \we \omega_{\nu ]}
\eeq 
using the techniques of \cite{romp98,bial96}. However, we leave 
the details of this construction beyond the scope of the present paper. 

According to the prescriptions of precanonical quantization 
\cite{qs96,bial97,lodz98,ijtp2001,opava2001,pla2001}   
we set 
\beq
\hat{\pi}_a^{\nu\mu} 
= - i\hbar\varkappa \gamma^\mu \frac{\der}{\der A^a_{\nu}} , 
\eeq
where $\gamma^\mu$ are the generating elements of 
the space-time Clifford algebra. 
However, this expression is not consistent with the constraints  
(2.7). Let us take the latter into account as 
the constraints on the physical quantum states   
\beq
\hat{\pi}_a^{(\nu\mu)}\left |\Psi\right >{}^{phys} =0 ,   
\eeq
that implies $\left < \right .\hat{\pi}_a^{(\nu\mu)} \left . \right > =0$ 
in accordance with (2.7). 
%
From (2.4) we obtain the DW Hamiltonian operator 
\beq
\what{H} = 
 \frac{1}{2} \hbar^2\varkappa^2 \frac{\der}{\der A_a^\mu\der A^a_\mu } 
- \frac{1}{2}ig\hbar\varkappa  C^a{}_{bc}A^b_\mu A^c_\nu 
\gamma^\nu \frac{\der}{\der A^a_\mu } \; , 
\eeq
or symbolically,  
$$
\what{H} = \frac{1}{2} \hbar^2\varkappa^2 \der_{AA} 
- \frac{1}{2}ig\hbar\varkappa\,CA\!\not\hspace{-3.5pt} A\,\der_A.  
$$
The quantum states are represented by  
Clifford-valued wave functions $\Psi (A^\mu_{a}, x^\nu)$ 
\beq
\Psi = \psi + \psi_\mu\gamma^\mu 
+ \frac{1}{2!} \psi_{\mu\nu}\gamma^{\mu\nu} +... 
\eeq
which fulfill  
the covariant Schr\"odinger equation proposed in 
\cite{qs96,bial97,lodz98} 
\beq
i\hbar\varkappa \gamma^\mu\der_\mu \Psi =\what{H} \Psi . 
\eeq

Note that the issue of gauge invariance is not directly relevant at this 
stage because it is related to the sections $A_a^\nu (x)$ 
of the field bundle over space-time 
rather than to the  fiber coordinates $A_a^\nu$ 
of this bundle which appear in  (3.2), (3.4).

It should be empasized that the DW Hamiltonian operator 
of pure Yang-Mills theory, eq.  (3.4),  
is not a scalar quantity as e.g. in the case 
of scalar field theory 
\cite{qs96,bial96,lodz98,pla2001}.  
It can be decomposed into two parts: 
\beq 
\what{H} = \what{H}{}^0 + \what{H}{}^1, 
\eeq
where $\what{H}{}^0$ is the free scalar part 
and  $ \what{H}{}^1=: \what{H}{}_\mu \gamma^\mu$ 
is the matrix interaction part. 
The presence of the latter term 
necessitates the use of a general Clifford-valued wave function 
(3.5) instead of a  simple wave 
function of the type $\Psi = \psi + \psi_\nu\gamma^\nu$ 
which is sufficient 
in the theories with a scalar 
DW Hamiltonian operator.  

To see it let us write the covariant Schr\"odinger equation (3.6) 
componentwise assuming $\psi_{\mu\nu}=0$. 
Here we set for simplicity $\hbar\varkappa=1$. From (3.6) it follows 
\beqa
i \der_\mu \psi^\mu &=& \what{H}{}^0\psi + \what{H}{}_\mu\psi^\mu , \\
i \der_\mu \psi &=& \what{H}{}^0\psi_\mu + \what{H}{}_\mu\psi , \\ 
i\der_{[\mu} \psi_{\nu ]} &=&\what{H}{}_{[\mu}\psi_{\nu ]} .
\eeqa

If $\what{H}{}_\mu = 0$ and there are no external fields, i.e. 
$\der_\mu \what{H}{} = 0$,  then (3.10) is the integrability 
condition of (3.9).  
Consequently,  the higher antisymmetric components of $\Psi$ 
decouple and it is sufficient to consider the wave functions of the type  
$\Psi = \psi + \psi_\nu\gamma^\nu$. 

If   $\what{H}{}_\mu \neq 0$  and $\der_\mu  \what{H}{}= 0$ 
the integrability condition of (3.9) takes the form  
$$
\what{H}{}^0 i \der_{[\mu} \psi_{\nu ]} - \what{H}{}_{[\mu} \what{H}{}^0 
\psi_{\nu ]} 
- \what{H}{}_{[\mu} \what{H}{}_{\nu ]}\psi 
= 0,  
$$
i.e. 
\beq
\what{H}{}^0 (i \der_{[\mu} \psi_{\nu ]} - \what{H}{}_{[\mu}\psi_{\nu ]}  ) 
+ [\what{H}{}^0,\what{H}{}_{[\mu}] \psi_{\nu ]} - \half [\what{H}{}_{\mu}, 
\what{H}{}_{\nu}]\psi = 0 . 
\eeq 
Using the antisymmetry of $C_{abc}$ we can prove that 
\beq
[\what{H}{}^0,\what{H}{}_{\mu } ]  = 0 . 
\eeq
However, 
 the Jacobi identity and the antisymmetry of $C_{abc}$  yield  
\beq 
{}[\what{H}{}_\mu, \what{H}{}_\nu ] 
= C^e{}_{cf} C^d{}_{eb} A^b_l A^c_{[\mu} A^f_{\nu]}\frac{\der}{\der A^d_l}
\neq 0 .   
\eeq
 Therefore, (3.10) is no longer the integrability condition of (3.9) 
and the assumption $\psi_{\mu\nu}=0$ turn out to be too restrictive. 
Thus, in the case of 
Yang-Mills  theory 
the truncation to the lower components of $\Psi$ 
is not justified and the higher antisymmetric 
components of $\Psi$ need to be taken into account 
in general.

\newcommand{\scalarproduct}{ 
From the 
covariant Schr\"odinger equation and its complex (hermitian??)  
conjugate 
\beq
i\hbar\varkappa \gamma^\mu \der_\mu\Psib = - \what{H}\Psib \quad ??
\eeq 
one can derive the following conservation law: 
\beqa
\der_\mu \, tr \int\! dy \, 
(\Psib  \gamma^\mu \Psi) 
&=& tr \!\int\! dy \, \der_\mu \Psib \gamma^\mu \Psi 
+ \Psib \gamma^\mu \der_\mu \Psi 
\nn \\
&=& - \frac{1}{i\hbar\kappa} tr\! \int\! dy \, 
(\what{H}\Psib)\Psi -\Psib (\what{H}\Psi) 
\;\;=\;\; 0  \quad ?? 
\eeqa 
provided the DW Hamiltonian is hermitian with respect to 
the scalar product 
\beq
(\Phi,\Psi):= tr \int\! dA \, \Phib \Psi ,   
\eeq
where $dA = \Pi_{\mu}\Pi_a \ dA^a_\mu$. 

... 

what is $\Phib$ exactly? complex conjugate and??? 

...

Ehrenfest theorem  ... 

...

gauge conditions ... 

... 

\subsection{The temporal gauge}

The need of gauge fixing??

In the temporal gauge $A^0_a=0$ the DW Hamiltonian takes the form 
\beq
\what{H} = 
 - \frac{1}{2} \hbar^2\varkappa^2 \frac{\der}{\der A_{ai}\der A_{ai} } 
- \frac{ig}{2}\hbar\varkappa  C^a{}_{bc}A^b_i A^c_j 
\gamma^j \frac{\der}{\der A^a_i } . 
\eeq
In this gauge the Gauss law has to be imposed as a condition on 
the physical states: 
\beq
\der_i< \what{\pi}{}_a^{0i}> 
+ g <C_{abc}A_{bi}\what{\pi}{}_c^{0i}> = 0 .  
\eeq

... 

... depends on the discussion of the Ehrenfest theorem! 

...

CHECK if the Gauss law constraint commutes with ${\cal E}$. 

} 

\section{A relation with the functional Schr\"odinger representation} 

In this section we explore how the seemingly unusual precanonical 
quantization of Yang-Mills theory is related to the familiar canonical 
quantization in the functional Schr\"odinger representation.

\subsection{Canonical quantization of pure Yang-Mills theory. A reminder} 

Let us briefly recall the functional Schr\"odinger representation 
of Yang-Mills theory in the temporal gauge $A^a_0(x)=0$ 
\cite{feynm,hatf85,luscher,rossi,hatfield,mansfield}.  
The canonical momenta  are given by 
\beq
p^i_a (\bx) := \frac{\delta \int\!d\bx\, L}{\delta (\der_t A^a_i) (\bx)}= 
F_a^{i0} (\bx) , 
\eeq
and the canonical Hamiltonian functional 
(in the metric signature $+-...-$) is  
\beqa
{\BH}[A^a_i(\bx), p^i_a (\bx)] 
&:=& \int \! d\bx\, 
\left ( p^i_a (\bx) \der_{_t} A^a_i (\bx) - L \right ) 
\nn \\
&=& \int \! d\bx\, 
\left ( \half p^i_a(\bx) p^i_a(\bx) 
+ \frac{1}{4}  F_a{}^{ij}(\bx) F_a{}^{ij}(\bx)  \right ) . 
\eeqa 
Henceforth the bold capital letters denote functionals and 
the small bold letters denote the spatial components of 
space-time vectors, e.g. $x^\mu=:(\bx, t)$; we also set $\hbar=1$.

The canonical momenta 
are represented (in the $A(\bx)$-representation) by the operators  
\beq
p^i_a (\bx) = - i  \frac{\delta}{\delta A^a_i(\bx)} .
\eeq  
The quantum states are given by the wave functionals 
$\BPsi= \BPsi([A^a_i(\bx)],t)$ which fulfill the 
 functional differential Schr\"odinger equation 
\beq
i\der_t \BPsi = 
 \int \! d\bx\, 
 \left ( - \half\ \frac{\delta}{\delta A^i_a(\bx)}\frac{\delta}{\delta A^i_a(\bx)} 
+ \frac{1}{4} F_a{}^{ij}(\bx) F_a{}^{ij}(\bx)  
 \right) \BPsi .
\eeq

The necessity for regularization arises 
here 
 because of the second variational derivative at equal points. 
We introduce a point-splitting based on a regulator 
$K_\epsilon(\bx,\bx')$ satisfying 
$$
\lim_{\epsilon\rightarrow 0} K_\epsilon(\bx,\bx') = \delta (\bx-\bx') .
$$
The regularized functional Laplacian in (4.4) takes the form 
\beq
\int \! d\bx \int \! d\bx'\,
K_\epsilon(\bx,\bx')
\frac{\delta}{\delta A^i_a(\bx)}\frac{\delta}{\delta A^i_a(\bx')} \; . 
\eeq



In addition to the Schr\"odinger equation the wave functional 
fulfills the Gauss law constraint on the physical states:   
\beq
\left ( \der_i \frac{\delta}{\delta A^a_i(\bx)} \, + \,  
g C^a{}_{bc} A^b_i \frac{\delta}{\delta A^c_i(\bx)} 
\right ) \BPsi = 0 ,  
\eeq
which implies that the physical wave functionals are gauge 
invariant under  ``small'' (i.e. topologically trivial) 
time-independent gauge transformations.  


\subsection{The derivation of the functional Schr\"odinger 
representation from the precanonical approach} 

It is definitely of interest to understand how the functional 
Schr\"odinger representation of quantum Yang-Mills theory  
can be connected with 
precanonical quantization. In what follows 
we will show that the functional differential 
Schr\"odinger equation (4.4) can be derived from the 
covariant Schr\"odinger equation of the precanonical approach, eq. (3.6). 
The relation of this kind has been studied in the case of scalar 
field theory in \cite{pla2001}. 
Its extension to the pure Yang-Mills fields in the temporal gauge 
is presented below. 

The basic idea is that the Schr\"odinger wave functional 
$\BPsi([A (\bx)],t)$, which represents the probability amplitude of 
simultaneously observing the field {\em configuration} 
$A=A (\bx)$ at the moment of time $t$,  can be seen as a joint probability
amplitude of 
simultaneously observing the respective {\em values} $A (\bx)$ at the  
spatial points $\bx$ (at the moment of time $t$). 
The amplitude of observing the value $A (\bx)$ at the  
spatial point $\bx$  
is 
given by the wave function $\Psi (A,\bx,t)$ restricted to the 
Cauchy surface $\Sigma$: $(A_\mu^a=A_\mu^a(\bx), \; t=const)$ 
    and taken at the point $\bx$. 
Using the result of \cite{pla2001},  in the temporal gauge 
$A_0^a(\bx)=0$ 
the corresponding composed amplitude is written as 
\beq
\BPsi([A^a_i(\bx)],t) = 
tr\left \{ (1+\beta) e^{\varkappa \int d\bx \ln \Psi_\Sigma (A_i^a(\bx),\bx,t)}
\right \} =: tr\left \{  ||\BPsi ||\right \},  
\eeq 
where $\Psi_\Sigma (A(\bx),\bx,t)$ denotes the restriction of the 
Clifford-valued wave function $\Psi (A,x)$ to the Cauchy surface $\Sigma$.
This Ansatz establishes a link between 
the Clifford-valued wave function appearing in precanonical  quantization 
and the Schr\"odinger wave functional resulting from canonical quantization. 
Note that (4.7) can be expressed in terms of the product integration 
\cite{pi}.  


Let us show that using the Ansatz (4.7) we can derive the 
familiar functional differential Schr\"odinger equation 
of Yang-Mills theory from the covariant Schr\"odinger equation (3.6). 
For this aid let us find the Schr\"odinger-type 
equation fulfilled by the functional amplitude (4.7) 
taking into account the covariant Schr\"odinger equation 
obeyed by $\Psi$. 
From (4.7) we obtain
\beqa
i\der_t \BPsi &=& tr \left \{ ||\BPsi||\varkappa 
\int \!d\bx\, \Psi_\Sigma^{-1} i\der_t \Psi_\Sigma   
\right \} , 
\\ 
\frac{\delta\BPsi}{\delta A^i_a(\bx)}  &=&
tr \left \{   ||\BPsi|| \varkappa \Psi_\Sigma^{-1}  \frac{\der}{\der A_a^i} \Psi_\Sigma   
\right \} , 
\eeqa
\beqa
\frac{\delta^2\BPsi  }{\delta A_i^a(\bx)\delta A_i^a(\bx)}  
 &=& tr \left \{   ||\BPsi|| \left (
\varkappa^2 \Psi_\Sigma^{-1} \frac{\der}{\der A_a^i} \Psi_\Sigma \Psi_\Sigma^{-1} \frac{\der}{\der A_a^i} \Psi_\Sigma  
\right . \right . \nn \\
&& \hspace*{-65pt}\left . \left .  
- \varkappa\delta^{n-1} (0) \Psi_\Sigma^{-1} \frac{\der}{\der A_a^i} \Psi_\Sigma \Psi_\Sigma^{-1} \frac{\der}{\der A_a^i} \Psi_\Sigma  
  + \varkappa\delta^{n-1} (0) \Psi_\Sigma^{-1} 
\frac{\der^2}{\der A_a^i\der A_a^i } \Psi_\Sigma \right ) 
\right \} . 
\eeqa 
The singularity $\delta^{n-1} (0)$ ($n$ is the space-time dimension) 
arises from 
the second functional differentiation at equal points. 
The simplest regularization 
$$
K_\epsilon (\bx, \bx'):= \left\{ 
\begin{array}{ccl} 
1/\epsilon^{n-1} &\mbox{\rm if}& |\bx-\bx'| \leq \epsilon, \\
0 &\mbox{\rm if}& |\bx-\bx'| > \epsilon 
\end{array}
\right . 
$$
amounts to the replacement of 
$\delta (0)$ with the momentum space cutoff $1/\epsilon$. 
The latter has its counterpart in precanonical quantization 
as the constant $\varkappa$ which has the meaning of 
$1/\epsilon^{n-1}$. 
 Under the regularization 
$\delta^{n-1} (0) \rightarrow \varkappa$ we obtain 
\beq
 \frac{\delta^2\BPsi  }{\delta A_i^a(\bx)\delta A_i^a(\bx)}  
 = tr \left \{   ||\BPsi|| \, \varkappa^2 \Psi_\Sigma^{-1}
\frac{\der^2}{\der A^a_i \der A^a_i} \Psi_\Sigma 
\right \} . 
\eeq 

Now, let us substitute into (4.8) the expression of $i\der_t\Psi_\Sigma$ 
which one obtains from the wave equation on the restricted 
wave function $\Psi_\Sigma$. The latter is derived from the 
covariant Schr\"odinger equation (3.6):  
in the temporal gauge (i.e. by just dropping out $A^a_0$) 
we obtain 
\beqa
i\der_t \Psi_\Sigma = -i\alpha^i \left (\frac{d}{dx^i} 
- \der_{[i} A{}^a_{j]} (\bx) \frac{\der}{\der A^a_j } \right ) \Psi_\Sigma 
+ \beta \what{H} \Psi_\Sigma  , 
\eeqa 
where 
$$
\frac{d}{dx^i} : = \der_i + \der_i A^a_j (\bx)\frac{\der}{\der A^a_j } 
$$ 
is the total spatial derivative,  
and 
\beq
\what{H} = 
- \frac{1}{2} \hbar^2\varkappa^2 
\frac{\der^2}{\der A^a_{i}\der A^a_{i} } 
-  \frac{ig}{2}\hbar\varkappa  C^a{}_{bc}A^b_i A^c_j  
\gamma^j  \frac{\der}{\der A^a_i  } 
\eeq
is the DW Hamiltonian operator in the temporal gauge, cf. (3.4).    
The antisymmetrization in $\der_{[i} A^a_{j]} (\bx)$ in (4.12) 
is related to the fact that the symmetric part of the 
polymomentum operator $\sim \gamma^i \der / \der A^a_j$ 
vanishes on the physical (restricted) wave functions 
due to the quantum version of the constraints (2.7), eq.~(3.3).  
In fact, due to the presence of the projector 
$\half(1+\beta)$ in the definition of $||\BPsi||$ 
we actually use here 
a weaker version of (3.3): 
$$
\gamma^{(i} \frac{\der}{\der A^a_{j)}} \Psi_\Sigma (1+\beta) = 0.
$$


Thus, by substituting $i\der_t\Psi_\Sigma$ from (4.12) to (4.8),  
discarding the total divergence term 
$\int \!d\bx\, \Psi_\Sigma^{-1} \alpha^i \frac{d}{dx^i} \Psi_\Sigma$,  
and using (4.11), (4.13) we obtain 
\beq
  i\der_t \, tr \left \{ \phantom{\what{\BH}||}\hspace{-15pt} 
||\BPsi|| \right \} 
= tr \left \{ || \what{\BH}||  \, ||\BPsi|| \right \} , 
\eeq
where the matrix-valued Hamiltonian operator is 
\beqa
||\what{\BH}|| &=& \int d\bx \left ( 
-  \half \frac{\delta^2  }{\delta A_i^a(\bx)\delta A_i^a(\bx)}  
- \frac{ig}{2}  C^a{}_{bc}A^b_i(\bx) A^c_j(\bx)  
\gamma^j  \frac{\delta}{\delta A^a_i(\bx)}  
\right . \nn \\ 
&&  \qquad \qquad
+  i\der_{{[i}} A^a_{j]} (\bx) \gamma^i \frac{\delta }{\delta A^a_j(\bx)} 
\left . \phantom{\frac{1}{1}}\hspace{-9pt} \right ) . 
\eeqa

In order to understand the relation of the matrix Hamiltonian 
$||\what{\BH}||$ with the Schr\"odinger picture Hamiltonian (4.4)
let us consider a unitary transformation of $||\what{\BH}||$ 
\beq
\what{{\BH}}'  = e^{-i\BN}||\what{{\BH}}||e^{i\BN} , 
\eeq 
where $\BN$ is a functional operator. 
By a straightforward calculation we obtain 
\beqa
e^{-i\BN}||\what{{\BH}}|| e^{i\BN}
&=&  \half \int \! d\bx \, 
\left (-\frac{\delta^2 i\BN }{\delta^2 A (\bx)} 
- \left ( \frac{\delta i\BN}{\delta A (\bx)} \right )^ 2 
- 2 \frac{\delta i\BN}{\delta A (\bx)}\frac{\delta }{\delta A (\bx)} 
- \frac{\delta^2  }{\delta^2 A (\bx)} 
\right.
\nn \\ 
&& \left.
\hspace*{-75pt} 
+ \; 
2 e^{-i\BN} 
\left [- \frac{ig}{2}  C^a{}_{bc}A^b_j(\bx) A^c_i(\bx)  
+  i\der_{[i} A^a_{j]} (\bx) \right ] \gamma^i e^{i\BN}  
 \left ( \frac{\delta i\BN}{\delta A^a_j (\bx)} 
+ \frac{\delta }{\delta A^a_j (\bx)}\right ) 
\right ) . 
\eeqa

The condition that the transformed Hamiltonian $\what{{\BH}}'$ 
contains no 
terms with the first order functional derivatives yields 
\beq
\frac{\delta \BN}{\delta A_a^j(\bx)} = \gamma^i ( 
 \der_{[i} A^a_{j]} (\bx)
- \frac{g}{2} C^a{}_{bc}A^b_{j}(\bx) A^c_{i}(\bx) ) 
= \half \gamma^i F^a_{ij}(\bx) , 
\eeq
whence it follows 
\beq
\frac{\delta^2 \BN }{\delta^2 A (\bx)} 
\sim \gamma^i \der_i \,\delta (0) = 0 , 
\eeq 
%
%
\beq
\left ( \frac{\delta \BN}{\delta A (\bx)} \right )^2 = 
- \frac{1}{4} F^a_{ij} F^a_{ij} .  
\eeq

Therefore, with the aid of the transformation 
(4.16), where $\BN$ is a solution of (4.18), the matrix-valued 
Hamiltonian $||\what{\BH}||$ is transformed to the Schr\"odinger 
picture Hamiltonian operator of Yang-Mills theory: 
\beq 
\what{\BH}_S =  \int \! d\bx\, 
 \left ( -\half 
\frac{\delta}{\delta A^i_a(\bx)}\frac{\delta}{\delta A^i_a(\bx)} 
+ \frac{1}{4} F^a_{ij}(\bx) F^a_{ij}(\bx)  
 \right) .
\eeq
Correspondingly, the Schr\"odinger picture wave 
functional is given in terms of the Ansatz (4.7) as follows: 
\beq
\BPsi_S = tr \left \{ e^{-i\BN} ||\BPsi|| \right \} . 
\eeq 

Now, let recall that by eliminating $A_0^a$ in (4.12) 
we actually lost the information of the $\nu=0$ component of the 
DW Hamiltonian equations (2.5), which is the Gauss law. In order to 
restore it,  we have to require that $\BPsi_S$ is invariant 
under the gauge transformations of $A^a_i(\bx)$. As we have already 
noticed, this automatically leads to the Gauss law constraint (4.6). 

Thus, the total set of equations of the functional 
Schr\"odinger representation of pure Yang-Mills theory in the temporal gauge 
is derived  from the precanonical approach. 


Let us note that eq. (4.18) is formal because the second 
functional differentiation in (4.15) is formal and requires 
 a 
regularization. 
A regulator function 
$K_\epsilon(\bx,\bx')$ will appear then in the right hand side 
of (4.18) as 
$\int\! d\bx'\, K_\epsilon(\bx,\bx')\, \delta \BN/\delta A(\bx')$, 
thus making the transformation operator $\BN$ explicitly depending 
on the regulator. In the unregularized case $K_\epsilon(\bx,\bx') = 
\delta(\bx-\bx')$ there are no solutions to (4.18) 
in the class 
of single-valued functionals of one argument 
$A^a_i(\bx)$.\footnote{In \cite{pla2001} this 
issue was formally treated in the case 
of scalar field theory by resorting to the solution for $\BN$ in terms 
of a functional of two arguments. The present treatment which 
recalls the necessity of regularization seems to be more satisfactory.}

\newcommand{\integrability}{
We still have to check if () has solutions. The integrability check 
obtained by functional differentiation of both sides of () 
with respect to $A_d^k(\bx')$ and subsequent antisymmetrization 
 yields: 
\beqa
&&\frac{\delta }{\delta A_d^k(\bx')}\frac{\delta \BN}{\delta A_a^j(\bx)} 
- \frac{\delta }{\delta A_a^j(\bx)} \frac{\delta\BN}{\delta A_d^k(\bx')}
= 
\gamma^i 
\left (\frac{\der}{\der x^{[i}} \delta(\bx-\bx') 
- \frac{\der}{\der x'{}^{[i}} \delta (\bx'-\bx)\right ) 
\delta_{j]k} \delta^{ad} \nn \\ 
&&... ???? 
\eeqa

...

\beq
\BN [A(\bx), F(\bx)]
= \half \int\!d\bx \gamma^i F^a_{ij}(\bx) A_a^j(\bx) ?????? 
\eeq
where the field strength variables $F^a_{ij}$ are considered 
as independent from the potentials $A_a^j(\bx)$. 

gauge condition ... Gauss law ..... 

... ?????? 

\beq
\int\!d\bx 
\left ( \der_i \frac{\delta}{\delta A_{ai}}   
+? g C^a_{bc} A^b_i \frac{\delta}{\delta A_{ci}}
\right ) \BPsi
\eeq

How to implement gauge conditions precanonically?????? 

} 

\section{On the spectrum of pure Yang-Mills theory}

The spectrum of a DW Hamiltonian 
divided by $\varkappa$ is related to the 
mass spectrum of the corresponding theory \cite{bial97,lodz98} 
 (prior to renormalization which is supposed  
to remove   $\varkappa$ from the physical 
results of the theory \cite{inprep}). 
For example, for a free scalar field the  DW Hamiltonian operator 
$$\what{H} = -\half\hbar^2\varkappa^2\der_{\phi\phi} 
+\half \frac{m^2}{\hbar^2}\phi^2$$ 
has a discrete spectrum 
$$ \chi_N = \varkappa m(N+\half).$$ 
The physical particles are related to the transitions 
between these stationary states.  Those are only possible for 
$\Delta N = \pm 1$.  Thus, the mass of a physical particle in this 
theory equals to $m$. 
This example demonstrates that the spectrum 
of the DW Hamiltonian operator of the Yang-Mills field 
contains information about 
the mass gap in quantum Yang-Mills theory.  

The spectrum of the DW Hamiltonian (3.4) is not easy to analyze because 
of the matrix interaction term 
\mbox{$CA\!\not\hspace{-3.5pt}A\,\der_A$}. 
In this section we show 
that there exists a 
formal 
transformation of $\what{H}$: 
\beq 
\what{H}' = e^{-i\hat{N}}\what{H}e^{i\hat{N}} 
\eeq 
 such that the term $CA\!\not\hspace{-3.5pt} A\,\der_A$ 
is transformed to a more familiar scalar potential term. 

For the transformed DW Hamiltonian we obtain 
(in the symbolic notation, for short)
\beqa
\what{H}' 
&=&  \frac{\hbar^2\varkappa^2}{2}\der_{AA}
+ \frac{\hbar^2\varkappa^2}{2}\der_{AA} (iN) 
+ \frac{\hbar^2\varkappa^2}{2} \der_A (iN) \der_A (iN) 
+ \hbar^2\varkappa^2 \der_A (iN) \der_A 
\nn \\
&&- e^{-i\hat{N}}( \frac{ig \hbar\varkappa }{2} 
CA\!\not\hspace{-3.5pt} A) \der_A (iN)e^{i\hat{N}} 
 - e^{-i\hat{N}} (\frac{ig \hbar\varkappa }{2} 
CA\!\not\hspace{-3.5pt} A)e^{i\hat{N}} 
\der_A . 
\eeqa 
If we wish to transform away the $\der_A$ terms from the 
DW Hamiltonian, the operator $\hat{N}$ has to fulfill the equation
\beq
\hbar\varkappa\frac{\der}{\der A_a^\mu} N = 
\half g C^a{}_{bc} A^b_\mu A^c_\sigma  \gamma^\sigma , 
\eeq
whence it follows 
\beqa
\der_A (iN) \der_A (iN) &=& -\frac{1}{4}g^2 CAACAA , 
\nn \\  
igCA\!\not\hspace{-3.5pt} A\,\der_A (iN) &=& - \frac{1}{2} g^2CAACAA , 
\\ 
\der_{AA} N &=& 0,  \nn  
\eeqa 
the latter being a consequence of the total antisymmetry 
of $C_{abc}$. 
Then the transformed Hamiltonian takes the form  
\beqa
\what{H}' 
 &=& \frac{1}{2}\hbar^2\varkappa^2 \frac{\der^2}{\der A_a^\mu\der A^a_\mu } 
+ \frac{1}{8}g^2 
C^a{}_{bc}A^b_\mu A^c_\nu C_{ade}A^d{}^\mu A^e{}^\nu  . 
\eeqa

In the temporal gauge $A_0^a=0$ 
the transformed DW Hamiltonian assumes the form 
\beq
\what{H}'= 
- \frac{1}{2}\hbar^2\varkappa^2  
\frac{\der}{\der A^a_i\der A^a_i } 
+ \frac{1}{8}g^2 
C^a{}_{bc} C_{ade} A^b_i A^c_j A^d_i A^e_j  .  
\eeq
Note that it admits a factorization 
\beq
\what{H}'= - \frac{1}{2} Q^\dag{}^a_i Q^a_i   
\eeq 
with 
\beq
Q^a_i =  i \hbar\varkappa\frac{\der}{\der A^a_i} 
- \frac{1}{2}g  C^a_{bc}  A^b_i A^c_j \gamma^j ,   
 \nn
\eeq 
that might be a starting point of an analytic study based 
on the multidimensional Darboux transformation technique 
\cite{kamran}. 

The operator (5.6) is known to have  purely discrete spectrum 
bounded from below \cite{simon}. 
Hence, the discreteness of the spectrum of $\what{H}'$ 
together with the connection between 
precanonical quantization of Yang-Mills theory and 
 the standart 
canonical quantization in the functional Schr\"odinger representation 
may point to the existence of the mass gap in 
quantum pure Yang-Mills theory. 

However, the problem with this argumentation  
is that the right hand side of (5.3) 
is not a gradient in 
the $A$-space. 
As a result the solution of (5.3) can be understood only as a functional 
\beq
N (A, [{\cal C}])= \half \frac{g}{\hbar\varkappa}
\int_{{\cal C}_{[A_0, A]}} C^a{}_{bc} 
A^b_\mu A^c_\sigma \gamma^\sigma d A_a^\mu 
\eeq 
which depends on a path 
${\cal C}_{[A_0, A]}$ in the $A$-space 
connecting an arbitrary ``initial'' point $A_0$ with the point $A$. 
The corresponding transformation (5.1) with 
a 
non-single-valued path-dependent $N$ given by (5.9) 
does {\em not\/} preserve the spectrum.

This situation is similar to the case of the Schr\"odinger operator 
with the magnetic field, $\what{H}_{V,{\bf A}}$,  
which can be transformed 
to a purely electric Schr\"odinger operator $\what{H}_{V,{\bf 0}}$ 
using a path-dependent transformation 
similar to (5.1) with $N\sim \int_C {\bf A}(\bx) \cdot d\bx$ 
\cite{dirac}, the spectrum of $\what{H}_{V,{\bf 0}}$ being 
different from the spectrum of  $\what{H}_{V,{\bf A}}$. 
Nevertheless, though this transformation does not preserve the spectrum, 
the presence of the magnetic field, 
i.e. the availability of the path-dependent transformation, 
in general is known to 
promote  
the discreteness of the spectrum:  if $\what{H}_{V,{\bf 0}}$ has 
a discrete spectrum then the same is true for $\what{H}_{V,{\bf A}}$ 
whatever the magnetic potential ${\bf A}(\bx)$ \cite{shubin}. 

Now, let us note that the original DW Hamiltonian in the temporal gauge, 
eq. (4.13),  can be rewritten in the form of 
the Schr\"odinger operator with both an ``electric'' ${\cal U} (A)$ 
and a ``magnetic''  ${\cal A} (A)$ field in the $A$-space: 
\beq
\what{H} = \what{H}_{{\cal U},{\cal A}} 
:= \half (-i\hbar\varkappa \der_A + {\cal A} (A))^2 + {\cal U} (A) , 
\eeq
where ${\cal A} (A) 
:= \half g \mbox{$CA\hspace{-1.5pt} \not \hspace{-4.0pt} A$} $ 
and ${\cal U} (A) := \frac{1}{8}g^2 CAA\,CAA$. 
The transformation (5.1), (5.9)  to the purely 
``electric'' Hamiltonian (5.5) just transfers 
the contribution of the Clifford-valued 
``magnetic'' field ${\cal A} (A)$  in (5.10) 
to the path-dependent factor $e^{iN}$. 
By analogy with the properties of the standard magnetic Schr\"odinger 
operator, it could be expected that the discreteness of the 
spectrum of (5.6) is a sufficient condition for the 
discreteness of the spectrum of (5.10) and, therefore, (4.13). 

However, this statement essentially relies on the extrapolation of the 
relevant theorems about the standart magnetic Schr\"odinger operator, 
which are proven using the properties of the Hilbert space of 
complex-valued functions,  to the 
operator  (5.10) with a Clifford-valued ``magnetic'' field,  
whose study should necessarily involve a corresponding 
Hilbert space theory of Clifford-valued functions. 
The validity of this extrapolation is not obvious and requires 
a separate study 
which needs  
a well established functional analysis of Clifford-valued 
functions and operators. Note that in pseudo-euclidean space the 
latter can be problematic because the natural scalar product 
\beq 
\left < \Psi, \Psi \right > = \int\! [dA]\, \left ( 
\psib\psi + \psib_\mu \psi^\mu 
+ \psib_{\mu\nu} \psi^{\mu\nu} + ... \right), 
\eeq 
where $[dA]$ is a measure in the space of Yang-Mills potentials,  
is not positive definite.

\medskip 

In conclusion, let us note that the precanonical framework 
seems to represent a foundation of 
field quantization that is better defined mathematically 
than  canonical or path integral quantization, which 
are known to require ad hoc regularizations and 
involve mathematical 
constructions whose rigorous definition is often problematic.  
%
%
We have essentially 
demonstrated that the (unregularized) functional Schr\"odinger 
representation resulting from the standard canonical quantization 
is a singular limit $\varkappa \rightarrow \delta^{n-1}(0)$
of precanonical quantization.  
In addition, precanonical quantization enabled us to 
relate the 
mass gap problem for quantum 
Yang-Mills theory to the spectral problem for 
the magnetic Schr\"odinger operator  
with a Clifford-valued ``magnetic'' field 
in the space of Yang-Mills potentials.

\bigskip 

{\bf Acknowledgements: } I express my gratitude to Prof. R. Schrader for 
his kind hospitality at the FU Berlin and a partial financial support 
and to Dr. S. Davis for his useful remarks concerning the text of 
the manuscript. I also thank Prof. A. Wipf, Dr. T. Strobl, Prof. R. Picken 
and Prof. S.~B. Leble for useful discussions during my visits to their 
Institutes.


\renewcommand{\baselinestretch}{0.90} 
\setlength{\parskip}{-0.0pt}

\end{document}